\documentclass[a4paper, 12pt]{article}
\usepackage{jheppub}
\usepackage{amssymb} 
\usepackage{amsmath}
\usepackage{mathtools}
\usepackage{amsfonts}    
\usepackage{dsfont}
\usepackage{young}
\usepackage[vcentermath]{youngtab}
\usepackage[mathscr]{euscript}
\usepackage{tensor}
\usepackage{xcolor}

\newcommand{\be}{ \begin{equation}}
\newcommand{\ee}{\end{equation}}
\newcommand{\bea}[1]{\begin{eqnarray}\label{#1} }
\newcommand{\eea}{\end{eqnarray}}

\def\ZZZ{{\hskip-3pt\hbox{ Z\kern-1.6mm Z}}}

\def\zzz{{\hskip-3pt\hbox{ z\kern-1mm z}}}

\def\one{{\hbox{ 1\kern-.8mm l}}}
\def\zero{{\hbox{ 0\kern-1.5mm 0}}}

\allowdisplaybreaks

\title{Structure of the ${\cal N}=4$ chiral algebra}

\author[a,b]{Matthias R.\ Gaberdiel}
\author[c,d]{and Wei Li}
\affiliation[a]{Institut f\"ur Theoretische Physik,
ETH Z\"urich,\\
Wolfgang-Pauli-Strasse 27,
8093 Z\"urich, Switzerland}
\affiliation[b]{Kavli Institute for Theoretical Sciences, University of Chinese Academy of Sciences,\\
Beijing 100190, China}
\affiliation[c]{Institute of Theoretical Physics, Chinese Academy of Sciences, \\
Zhongguancun East Road 55, 
Beijing 100190 , China}
\affiliation[d]{Peng Huanwu Center for Fundamental Theory, 
 Hefei, Anhui 230026 , China
}
\emailAdd{gaberdiel@itp.phys.ethz.ch}
\emailAdd{weili@mail.itp.ac.cn}

\abstract{The chiral algebra of 4D $\mathcal{N}=4$ SU$(N)$ super-Yang-Mills theory is an $\mathcal{N}=4$ superconformal vertex operator algebra. 
We analyse the structure of this algebra by studying recursively the constraints that are required by the associativity of the operator product expansion. 
We find that the algebra is uniquely characterized by the central charge (which can take an arbitrary value), without any additional free parameter. 
Furthermore, the truncation pattern of the OPE coefficients suggests that the  algebra cannot 
arise from the symmetric orbifold.
}

\begin{document}

\maketitle

\section{Introduction}

Some time ago it was noted that one can associate a 2D chiral algebra (or VOA) to any 4D ${\cal N}=2$ superconformal theory \cite{Beem:2013sza}, which can also be obtained by applying the holomorphic-topological twist of \cite{Kapustin:2006hi} in the presence of the $\Omega$-background \cite{Oh:2019bgz,Jeong:2019pzg}.
The resulting algebras have been studied in some detail, see e.g.\ \cite{Beem:2017ooy,Arakawa:2023cki,Bonetti:2018fqz}, although a number of questions have not been answered up to now. 
For example, for the arguably most important case of 4D ${\cal N}=4$ SYM, the resulting 2D algebra has small ${\cal N}=4$ superconformal symmetry, and consists, in addition to the superconformal generators, of the BPS multiplets 
\be
\mathbb{V} = \bigl( {\cal N}=4 \bigr)  \oplus \bigoplus_{s} R^{(s)} \ ,
\ee
where $R^{(s)}$ is the ${\cal N}=4$ BPS multiplet generated by a field with conformal dimension and $\mathfrak{su}(2)$ spin equal to $h=j=s$. 
Depending on which gauge theory one considers, the parameter $s$ takes the values 
\begin{align}\label{SYMtrun}
\hbox{SYM based on {\rm SU}(N):} \qquad & s=\tfrac{3}{2},2,\ldots, \tfrac{N}{2} \ ,
\end{align}
and similarily for ${\rm SO}(N)$ or ${\rm U}(N)$. For integer $N$, these algebras have been constructed in terms of symplectic bosons and free fermions \cite{Beem:2013sza,Bonetti:2018fqz,Arakawa:2023cki}, and they have central charge (for the ${\rm SU}(N)$ case)
\be\label{cN}
c_N = - 3 (N^2-1) \ . 
\ee
However, relatively little is known about the general structure of these algebras. 
In particular, two questions remain open:
\begin{enumerate}
\item Do these algebras also exist for general (non-integer) values of $N$, or correspondingly arbitrary values of $c$?
\item Are these algebras uniquely characterized by the central charge, or do they depend on any additional parameters?
\end{enumerate}
\smallskip

Seemingly similar ${\cal N}=4$ superconformal algebras have appeared in the context of the AdS$_3$/CFT$_2$ higher spin duality \cite{Gaberdiel:2013vva}. In that setup the relevant VOA is generated by a large ${\cal N}=4$ superconformal algebra with spectrum\footnote{The large ${\cal N}=4$ superconformal algebra has $R$-symmetry $\mathfrak{su}(2) \oplus \mathfrak{su}(2)$, and the small ${\cal N}=4$ superconformal algebra can be obtained from it by a contraction limit.} 
\be\label{largeN4}
(\hbox{large ${\cal N}=4$}) \oplus \ \bigoplus_{s=1}^{\infty} \hat{R}^{(s)} \ , 
\ee
where $\hat{R}^{(s)}$ is the non-BPS highest weight representation with conformal dimension $h=s$ and vanishing $\mathfrak{su}(2)$ spins $j_1=j_2=0$. These algebras were studied systematically in \cite{Beccaria:2014jra}, and it was found that 
\begin{enumerate}
\item they exist for general values of the central charge, or rather, general values of the levels of the two $\mathfrak{su}(2)$ algebras. 
\item They are characterized by an additional parameter that describes the $3$-point coupling\footnote{In the notation of \cite{Beccaria:2014jra}, the relevant free parameter is $w_{70}$ defined in eq.~(3.15).}
\be
\langle \hat{R}^{(1)} \hat{R}^{(1)}  \hat{R}^{(1)}  \rangle \ .  
\ee
\end{enumerate}
One may therefore wonder whether the algebra $\mathbb{V}$ exhibits a similar structure, i.e.\ whether it depends  on some free parameters in addition to the central charge.
\medskip

Another motivation for studying the algebra $\mathbb{V}$ abstractly comes from the work of \cite{Li:2023tyx}, in which the spectrum of $\mathbb{V}$ was reproduced from a worldsheet perspective, using the proposed worldsheet theory for free ${\cal N}=4$ SYM of \cite{Gaberdiel:2021qbb,Gaberdiel:2021jrv}. 
In particular, the analysis of \cite{Li:2023tyx} showed that the algebra $\mathbb{V}$ arises from (the BPS part of) an ${\rm AdS}_3 \times {\rm S}^3$ subsector of ${\rm AdS}_5\times {\rm S}^5$, consistent with the expectations from the twisted holography of \cite{Costello:2018zrm}.\footnote{See also \cite{Bonetti:2016nma} for a direct localization analysis of the simplest ``twisted" truncation of AdS$_5\times {\rm S}^5$ supergravity,
for which the chiral algebra is reduced to the $\mathfrak{su}(2)_{\textrm{R}}$ affine Kac-Moody subalgebra.} 
The physical state conditions on this ${\rm AdS}_3 \times {\rm S}^3$ inherit those from ${\rm AdS}_5\times {\rm S}^5$, and its BPS spectrum consists of the states
\begin{equation}\label{AdS3spec}
\mathbb{V} = \bigoplus_{w=1}^{\infty}\, R^{(\frac{w+1}{2} )} \ ,
\end{equation}
where $w=1,2,\dots,\infty$ labels the different spectrally flowed sectors of the worldsheet theory. This therefore suggests that the algebra should also naturally appear in the AdS$_3$/CFT$_2$ duality. 
More specifically, it should come from the  ``compactification independent" (i.e.\ the $\mathbb{T}^4$ independent) part of string theory on ${\rm AdS}_3 \times {\rm S}^3 \times \mathbb{T}^4$ with pure R-R flux \cite{Li:2023tyx}.
\smallskip

In a similar vein, it was argued in \cite{Costello:2020jbh} that applying the twisting procedure of \cite{Costello:2018zrm} to  supergravity on ${\rm AdS}_3\times {\rm S}^3 \times \mathbb{T}^4$ leads to a VOA of the form\footnote{See \cite{Fernandez:2024tue} for the analysis of the  ${\rm AdS}_3\times {\rm S}^3 \times {\rm K3}$ case, with a similar conclusion.
}
\begin{equation}
\tilde{\mathbb{V}} = \bigl( {\cal N}=4 \bigr)  \oplus \, H^*(\mathbb{T}^4) \otimes   \bigoplus_{s}   R^{(s)} \ ,
\end{equation}
where $H^*(\mathbb{T}^4)$ is the cohomology of $\mathbb{T}^4$. From this perspective, reducing to the ``compactification independent"  part should amount to dividing out the cohomology of $\mathbb{T}^4$, i.e.\ $H^*(\mathbb{T}^4)$.  This then gives rise to a VOA that has the same spectrum as $\mathbb{V}$ above. 
This spectrum should be independent of which specific background one considers, i.e.\ whether the ${\rm AdS}_3$ factor is supported by NS-NS or R-R flux since the BPS spectrum is the same everywhere in moduli space. 

The setup in \cite{Li:2023tyx} corresponds to a string background with pure R-R flux, and thus the chiral algebra of ${\cal N}=4$ SYM in 4D should agree with the algebra structure for pure R-R flux. 
On the other hand, for the background with pure (minimal) NS-NS flux, the dual CFT has been shown to agree exactly with the symmetric orbifold theory of $\mathbb{T}^4$, ${\rm Sym}_M(\mathbb{T}^4)$ \cite{Eberhardt:2018ouy,Eberhardt:2019ywk}, and thus, at this point in moduli space, one should make contact with a symmetric orbifold construction. Actually, it was suggested in \cite{Costello:2020jbh} that it should be possible to obtain the chiral algebra $\tilde{\mathbb{V}} $ from the symmetric orbifold, although the detailed dictionary was not spelled out there.

 While it is therefore difficult to make a direct comparison, we can use that the symmetric orbifold also has a very characteristic truncation pattern similar to eq.~(\ref{SYMtrun}) above: for finite $M$, the symmetric orbifold ${\rm Sym}_M(\mathbb{T}^4)$ has $c=6M$, and its single-particle chiral BPS spectrum consists of the finitely many states
\be\label{symorbspec}
\bigoplus_{w=1}^{M} \Bigl( R^{(\frac{w-1}{2})} \oplus 2 \cdot R^{(\frac{w}{2})} \oplus R^{(\frac{w+1}{2} )} \Bigr) \ , 
\ee
where $w$ denotes the different single-cycle twisted sectors. The different BPS states for each $w$ are related to one another by the fermionic modes associated to $\mathbb{T}^4$, and the ``compactification independent" part of the spectrum should therefore simply be \cite{Gaberdiel:2021jrv,Li:2023tyx}
\be\label{cNSym}
c = 6M : \qquad \bigoplus_{w=1}^{M} \, R^{(\frac{w+1}{2} )} \ . 
\ee
This line of thought would then suggests that the algebra $\mathbb{V}$ should also exhibit this second truncation property. 

Note that it is, a priori, not clear whether the VOA structure of the two constructions is the same, i.e.\ whether the pure R-R algebra (that is inherited from ${\cal N}=4$ SYM in 4D) will agree, as an algebra, with the pure NS-NS version that comes from the symmetric orbifold. 
In particular, one might wonder whether, just as for the higher spin algebras reviewed above, the algebra $\mathbb{V}$ is not uniquely fixed by its spectrum, but has additional parameters, e.g.\ related to the ratio of NS-NS to R-R flux.\footnote{This parameter, if it exists, would not be visible from the 4D perspective, i.e.\ the putative deformed chiral algebra would not arise from 4D SYM.}
\bigskip

In view of these different considerations, it is therefore worthwhile to study the abstract structure of the VOA associated to $\mathbb{V}$, and this is what we shall be doing in this paper.\footnote{Similar methods were also employed in \cite{Bonetti:2018fqz}. More recently, the 6D chiral algebra (which is more similar to the higher spin algebras of eq.~(\ref{largeN4})) was also studied using related techniques in \cite{Woolley:2025qbd}. }
 The results we find are somewhat surprising: 
\begin{enumerate}
\item The algebra does not seem to possess any free parameter, but is uniquely characterized by the central charge, which can take an arbitrary value. 
It therefore behaves quite differently from the higher spin algebras of \cite{Gaberdiel:2013vva,Beccaria:2014jra}. 
\item While the algebra we construct exhibits beautifully (and non-trivially) the truncation pattern for $c=c_N$ of eq.~(\ref{cN}), this does not seem to be the case for \linebreak $c=6M$ with $M$ integer, see eq.~(\ref{cNSym}). In particular, this second property is quite puzzling, and we shall discuss possible explanations of it in the Conclusions, see Section~\ref{sec:concl}. 
\end{enumerate}
\smallskip

The paper is organized as follows. 
In Section~\ref{sec:spectrum} we describe the spectrum of the VOA $\mathbb{V}$ in some detail and fix our conventions, see also Appendix~\ref{app:N4}. The systematic analysis of the VOA structure is explained in Section~\ref{sec:VOA}, and we summarize the results we find from the associativity analysis in Section~\ref{sec:assoc}. The truncation patterns are studied in Section~\ref{sec:Vtrunc}, and we discuss the implications of our analysis in Section~\ref{sec:concl}. We have included one Appendix in which our conventions for the supermultiplets are spelled out in some detail. 
\smallskip

\noindent {\bf Note added:}  
While this paper was being finalized we were made aware by the authors of \cite{BM} that they have also studied the algebra $\mathbb{V}$ abstractly. 

\section{The spectrum of the VOA}\label{sec:spectrum}

Let us begin by describing the spectrum of the ${\cal N}=4$ VOA $\mathbb{V}$ of \cite{Beem:2013sza}. 
The VOA $\mathbb{V}$ contains the (small) ${\cal N}=4$ superconformal algebra whose generators are 
\begin{itemize}
\item $\mathfrak{su}(2)$ currents $K^a$ with $a\in\{3,\pm\}$ at level $k$,
\item four supercurrents of conformal dimension $h=\frac{3}{2}$ that we denote by $G^{\alpha\beta}$,
\item the stress energy tensor $L$ of conformal dimension $2$ and central charge $c=6k$.
\end{itemize}  
Our conventions for the ${\cal N}=4$ generators are spelled out in Appendix~\ref{app:N4}. 

The remaining generators of the VOA $\mathbb{V}$ sit in BPS multiplets of the ${\cal N}=4$ algebra. More specifically, let us denote by $R^{(s)}$ the ${\cal N}=4$ multiplet consisting of the fields
\be
\begin{array}{lc}
            & \ \ \ \ \bigl[h =s \ , \ \ j=s\bigr] \\[4pt]
R^{(s)} :      \qquad        & 2 \cdot \bigl[h =s+\frac{1}{2} \ , \ \ j=s - \frac{1}{2}\bigr] \\[4pt]
              & \ \ \ \ \phantom{\ ,} \bigl[h =s+1 \ , \ \ j=s - 1\bigr] \ , 
\end{array}
\ee
where $h$ denotes the conformal dimension and $j$ the $\mathfrak{su}(2)$ spin. One would expect that, for generic values of $c$ (resp.\ $k$), one can define a VOA whose generating fields consist of 
\be\label{VOA}
\boxed{\mathbb{V} = \bigl( {\cal N}=4 \bigr)  \oplus \bigoplus_{s=\frac{3}{2},2,\ldots} R^{(s)} \ . }
\ee

Our aim is to determine the structure of this VOA. In particular, by analogy with the large ${\cal N}=4$ ${\cal W}_\infty[\lambda]$ algebra, see e.g.\ \cite{Gaberdiel:2013vva,Beccaria:2014jra}, one may expect that the VOA is not uniquely characterized by the central charge $c$, but depends on an additional parameter $\lambda$ that controls one of the low-lying structure constants.\footnote{In these examples the additional multiplets are not BPS, but it is not obvious why this should modify the structure of the VOA.} However, as we shall see in this paper, no such free parameter seems to exist for the VOA $\mathbb{V}$. 
\smallskip

\section{The VOA structure of $\mathbb{V}$}\label{sec:VOA}

In this section we shall analyse the structure of the VOA $\mathbb{V}$. 

\subsection{Spin/statistic of fields}\label{sec:spinstatistics}
Before we begin with describing this analysis in some detail, there is one important general remark we should make. 

In order for the spin $s$ multiplet to appear in the chiral algebra, its self-OPE must contain the vacuum (so that the two-point function of $W^{(s)}$ with itself is non-zero). 
Given that $W^{(s)}$ has conformal dimension and spin equal to $s$, $h=j=s$, the leading term of the OPE of $W^{(s)}$ with itself has the structure
\be
W^{(s)}_m(z) W^{(s)}_n(w) \sim \frac{\langle 0,0|\!| s,m;s,n\rangle}{(z-w)^{2s}} \, {\bf 1} + \cdots \ , 
\ee
where $\langle 0,0|\!| s,m;s,n\rangle$ is the Clebsch-Gordan coefficient describing the tensor product decomposition  
\be
|s,m\rangle \otimes |s,n\rangle \rightarrow |0,0\rangle \ . 
\ee
(The fact that this Clebsch-Gordan coefficient appears is a consequence of the global $\mathfrak{su}(2)$ symmetry of the problem.) Since the Clebsch-Gordan coefficients satisfy\footnote{This is just to say that for integer spin $s$, the vacuum representation appears in the symmetric part of the tensor product of $s\otimes s$, whereas for half-integer spin, it appears in the anti-symmetric part of the tensor product.} 
\be
\langle 0,0|\!| s,m;s,n\rangle = (-1)^{2s} \, \langle 0,0|\!| s,n;s,m\rangle \ ,
\ee
we have
\begin{align}
W^{(s)}_n(w) W^{(s)}_m(z)   \sim \frac{\langle 0,0|\!| s,n;s,m\rangle}{(w-z)^{2s}} \, {\bf 1} + \cdots \sim
W^{(s)}_m(z) W^{(s)}_n(w) \ . 
\end{align}
As a consequence, all the $W^{(s)}_m$ fields are bosons, even for half-integer spin. 
Let us denote the different fields of each spin $s$ multiplet as 
\be
\begin{array}{lll}
W^{(s)}_{m}  \qquad & m=-s,\ldots,s  \qquad \qquad \qquad & (h=s)\\
W^{(s)[\alpha]}_{m}  \qquad & m=-(s-\tfrac{1}{2}),\ldots,(s-\tfrac{1}{2})  \qquad \qquad \qquad & (h=s+\tfrac{1}{2})\\
W^{(s)\uparrow}_{m}  \qquad & m=-(s-1), \ldots, (s-1)   & (h=s+1) \ , 
\end{array}
\ee
where $\alpha\in\{ \pm\}$. Then the bosonic fields are 
\be\label{bosons}
\hbox{bosons:} \qquad W^{(s)}_m \ , \ \ (|m|\leq s) \qquad W^{(s)\uparrow}_n \ , \ \ (|n|\leq s-1)\ ,  \qquad s = \tfrac{3}{2},2,\tfrac{5}{2},\ldots \ , 
\ee
while the fermions are 
\be
\hbox{fermions:} \qquad W^{(s)[\alpha]}_r \ , \ \ (|r|\leq s-\tfrac{1}{2})\ , \ \alpha\in\{\pm\} \ , \qquad s = \tfrac{3}{2},2,\tfrac{5}{2},\ldots \ . 
\ee

Our algebra therefore necessarily breaks the usual spin-statistics theorem, and as a consequence will be non-unitary. We should mention that the presence of the $\mathfrak{su}(2)$ symmetry is crucial for reaching this conclusion, and this is a first sign that the structure of $\mathbb{V}$ differs from that of the more familiar higher spin algebras of e.g. \cite{Gaberdiel:2013vva,Beccaria:2014jra}. 

\subsection{The structure of the associativity analysis}\label{sec:assoc}

In order to simplify the analysis, we shall in the following only consider the bosonic subalgebra of $\mathbb{V}$. As we will show, this is actually sufficient to answer the two questions we raised in the introduction. The bosonic subalgebra of $\mathbb{V}$ contains 
\begin{itemize}
\item the $\mathfrak{su}(2)$ currents $K^a$,
\item the stress energy tensor $L$,
\item  the fields in (\ref{bosons})
\item normal ordered even products of the fermionic generators of $\mathbb{V}$.
\end{itemize} 
In the following we shall give a schematic description of our ansatz, and describe what constraints we imposed and what conditions they give rise to. The actual computations were done using Mathematica, in particular, using the package  {\sf OPEdefs} developed by Kris Thielemans \cite{thelemansthesis,opedefs}.

The basic idea of the procedure is to make the most general ansatz for the various OPEs, and then to determine the different structure constants by imposing associativity. 
Given that the algebra is infinite-dimensional, there are in principle infinitely many OPEs to consider, and we will not be able to do so in closed form. 
Instead we shall study the problem recursively as follows.

\begin{enumerate}
\item Fix the OPEs of the bosonic ${\cal N}=4$ generators with the bosonic fields in the multiplet $W^{(s)}$ --- this can be done in closed form, and the results are described in detail in Appendix~\ref{app:A.1}. 
\item Make an ansatz for the OPE of $W^{(s_1)} \times W^{(s_2)}$. Generically, any field of spin less or equal to $s_1+s_2-1$ can appear on the right-hand-side of this OPE, so the number of terms in the ansatz grows with $s_1+s_2$, and hence this needs to be done case by case.\footnote{In general we will also need an ansatz for the OPEs where one or both of $W^{(s)}$ are replaced by $W^{(s)\uparrow}$. The same also applies to the various considerations below.} 
\item Fix the coefficients in the ansatz of the OPE $W^{(s_1)} \times W^{(s_2)}$ by checking the associativity with the bosonic ${\cal N}=4$ generators
\begin{equation}
(\textrm{bosonic }\mathcal{N}=4) \times W^{(s_1)} \times W^{(s_2)}     \ .
\end{equation}
This will fix some of the structure constants in the ansatz, but not all of them.
\item Determine the remaining structure constants by considering the associativity of 
\be\label{genasso}
W^{(s_1)} \times W^{(s_2)} \times W^{(s_3)} \ . 
\ee
\end{enumerate}

The problem exhibits a natural graded structure that arises as follows. 
In order to analyse the associativity of (\ref{genasso}) we need to have an ansatz for all the OPEs that appear in the calculation.
Note that they include not only the OPEs of any two of $W^{(s_i)} \times W^{(s_j)}$, but also the OPEs of $W^{(s_1)}$, say with any of the fields that appear in the OPE of $W^{(s_2)} \times W^{(s_3)}$. 
Given that the latter includes generically all fields of spin up to $s_2+s_3-1$, in order to study the associativity of (\ref{genasso}) with $s_1+s_2+s_3=P$, we need to make an ansatz for all OPEs of 
\be
W^{(s_1)} \times W^{(s_2)} \qquad \hbox{with $s_1 + s_2 \leq P-1$} \ . 
\ee
We will be able to perform this analysis for $P=\frac{9}{2}$ and $P=\frac{10}{2}=5$, see Section~\ref{sec:M92} and \ref{sec:M5}, respectively. As we shall see, already at these (low) levels, the analysis is quite complicated, and it will be difficult to go much beyond $P=5$. However, quite surprisingly, the constraints that arise from these low-lying considerations are already very restrictive. 

\subsection{The associativity analysis for grade $P=\frac{9}{2}$}\label{sec:M92}

At grade $P=\frac{9}{2}$, there is only one possible configuration for (\ref{genasso}), namely 
\be\label{firstassoc}
W^{(\frac{3}{2})} \times W^{(\frac{3}{2})} \times W^{(\frac{3}{2})} \ .
\ee
In order to analyse the associativity of \eqref{firstassoc}, we need to make an ansatz for 
\be
W^{(\frac{3}{2})}\times W^{(\frac{3}{2})} \qquad \hbox{and} \qquad 
W^{(\frac{3}{2})}\times W^{(2)} \ ,
\ee
i.e. we need to perform Step-$2$ and $3$ for each of these two OPEs in turn. 

\subsubsection*{The $W^{(\frac{3}{2})}\times W^{(\frac{3}{2})}$ OPE}
Since $W^{(\frac{3}{2})}_m$ has conformal dimension $h=\tfrac{3}{2}$, its OPE can only contain (in its singular part) fields of conformal dimension $0$, $1$ and $2$, and thus the most general ansatz is\footnote{For ease of notation we are suppressing here the $\mathfrak{su}(2)$ labels; their structure is fixed by the global $\mathfrak{su}(2)$ symmetry.}
\begin{align}\label{3232}
W^{(\frac{3}{2})} \times W^{(\frac{3}{2})} \sim & \, n_{\frac{3}{2}} \Bigl[ {\bf 1} + c_1 K + c_2 L + c_3 \partial K + c_4 :KK: \Bigr] 
 \oplus C_{\frac{3}{2} ,\frac{3}{2}}^{2} W^{(2)} \ , 
\end{align}
where we have grouped the fields according to the superconformal family in which they occur. (This is Step-2 for this case.) 

In order to perform Step-3 we now impose the associativity with the bosonic ${\cal N}=4$ generators. 
This can obviously not fix the overall normalization parameter $n_{\frac{3}{2}}$, nor can it constraint the structure constant $C_{\frac{3}{2}, \frac{3}{2}}^{2}$. 
However, it will determine the coefficients $c_i$ that describe the coupling to the other fields  in the vacuum representation of the ${\cal N}=4$ algebra (apart from the identity). 

With the natural conventions for the $\mathfrak{su}(2)$ structure constants, we find
\be
c_1 = \frac{3\sqrt{10}}{c} \ , \qquad c_2 = \frac{3 (c-3)}{c (9 + c)} \ , \qquad 
c_3 = \frac{3 \sqrt{5}}{\sqrt{2} c} \ . 
\ee
As for $c_4$, there are actually, a priori, three coefficients $c^{(j=0,1,2)}_4$, where $j$ labels the $\mathfrak{su}(2)$ spin $j$ of the resulting normal-ordered product. 
The term corresponding to $c_{4}^{(j=1)}$, i.e.\ the spin $j=1$ part of $(:KK:)_{j=1}$, can be rewritten in terms of $\partial K$, and thus we only need to consider $c^{(j=0,2)}_4$, for which we find
\be
c_4^{(0)} = - \frac{36 \sqrt{3}}{c (c+9)}  \ ,  \qquad c_4^{(2)} = \frac{18 \sqrt{6}}{c (c-6)} \ . \label{c4}
\ee
Note that for $c=6$, i.e.\ $k=1$, the second coefficient has a divergence, reflecting that the spin-$2$ combination of $:KK:$ is a null-field at $k=1$.

\subsubsection*{{The $W^{(\frac{3}{2})}\times W^{(2)}$ OPE}}
As anticipated, $W^{(2)}$ appears in the OPE of $W^{(\frac{3}{2})}\times W^{(\frac{3}{2})}$, and hence in order to study the associativity of (\ref{firstassoc}), we also need to make an ansatz for the OPE of $W^{(\frac{3}{2})}\times W^{(2)}$. 
This has the schematic form 
\begin{align}\label{322}
W^{(\frac{3}{2})} \times W^{(2)} \sim C_{\frac{3}{2} 2}^{\frac{3}{2}} \Bigl[ W^{(\frac{3}{2})} + d_1 :KW^{(\frac{3}{2})}: + d_2 \, \partial W^{(\frac{3}{2})}  \Bigr] 
\oplus C_{\frac{3}{2} ,2}^{\frac{5}{2}} \, W^{(\frac{5}{2})}  \oplus 
C_{\frac{3}{2} ,2}^{\frac{3}{2}\uparrow} \, W^{(\frac{3}{2})\uparrow}  \ . 
\end{align}
This is again Step-2 for this case. 
As regards Step-3, $d_1$ and $d_2$ are fixed by associativity with respect to the bosonic part of the ${\cal N}=4$ superconformal generators, 
\be
d_1^{(\frac{1}{2})} = - \frac{6 \sqrt{6}}{c+30} \ , \quad 
d_1^{(\frac{3}{2})} =  \frac{2 \sqrt{\frac{6}{5}}}{c-3} \ , \quad 
d_1^{(\frac{5}{2})} =  \frac{6 \sqrt{\frac{14}{5}}}{c-18} \ , \qquad
d_2 =   \frac{(c+3)}{3 (c-3)} \ ,
\ee
where, as before, $d_1^{(j)}$ denotes the coefficient in front of the spin $j$ part of the normal ordered product $:KW^{(\frac{3}{2})}:$. On the other hand, 
\be
C_{\frac{3}{2} ,2}^{\frac{3}{2}} \ , \quad C_{\frac{3}{2} ,2}^{\frac{5}{2}} \ , \quad C_{\frac{3}{2} ,2}^{\frac{3}{2}\uparrow}
\ee
are not determined by these conditions. (If we had imposed the full (supersymmetric) ${\cal N}=4$ constraints, the coefficient $C_{3/2, 2}^{3/2\uparrow}$ would have been fixed in terms of $C_{3/2, 2}^{3/2}$ since the two fields sit in the same ${\cal N}=4$ multiplet.)

\subsubsection*{The associativity of (\ref{firstassoc})}

With these preparations, we can now solve for the associativity of (\ref{firstassoc}), and we obtain
the condition:
\be\label{C32322}
\boxed{
C_{\frac{3}{2} ,\frac{3}{2}}^{2} \cdot C_{\frac{3}{2} ,2}^{\frac{3}{2}} = 
- 18 \sqrt{5}\,  \frac{(c+24) (c-3) }{c(9+c) (c-6)} \, n_{\frac{3}{2}}  \ . }
\ee
At this stage this is the only condition we find. 
We note that these structure constants diverge for $c=6$ and $c=-9$; this suggests that for these values of $c$ we need to set $n_{\frac{3}{2}}=0$, i.e.\ they describe the special cases where the ${\cal W}$ algebra truncates to the ${\cal N}=4$ algebra itself. We will comment on this issue below, see the discussion below eq.~(\ref{eq:ExN2}).

\subsection{The associativity analysis for grade $P=5$}\label{sec:M5}

At grade $P=5$, there is again only one possible configuration for (\ref{genasso}), namely 
\be\label{secondassoc}
W^{(\frac{3}{2})} \times W^{(\frac{3}{2})} \times W^{(2)} \ .
\ee
In order to analyse the associativity of \eqref{secondassoc}, we now need, in addition to the OPEs of eqs.~(\ref{3232}) and (\ref{322}), an ansatz for 
\be
W^{(2)}\times W^{(2)}\ ,  \qquad  W^{(\frac{3}{2})}\times W^{(\frac{5}{2})} \ , \quad \hbox{and} \quad 
W^{(\frac{3}{2})}\times W^{(\frac{3}{2})\uparrow} \ ,
\ee
i.e. we need to perform Step-2 and 3 for each of these OPEs in turn. 

\subsubsection*{The $W^{(2)}\times W^{(2)}$ OPE}
The most general ansatz for the $W^{(2)}\times W^{(2)}$ OPE is
\begin{align}
W^{(2)} \times W^{(2)} \sim &\, n_2  \Bigl[ {\bf 1} + K + L + \partial K + :KK: + :KKK: + :KL: + :K \partial K: + \partial L\Bigr] \nonumber  \\
& \oplus C_{2,2}^{2} \Bigl[ W^{(2)} + \partial W^{(2)} +  :KW^{(2)}: \Bigr]  \oplus C_{2,2}^{3} W^{(3)}   \\
& \oplus C_{2,2}^{2\uparrow}  W^{(2)\uparrow} \oplus C_{2,2}^{GG} :GG:  \oplus \, 
C_{2,2}^{:\frac{3}{2} \frac{3}{2}:} : W^{(\frac{3}{2})} W^{(\frac{3}{2})}: \ , \nonumber
\end{align}
where, to improve readability, we have not written out the coefficients in front of $K$, $L$, etc.\footnote{One may have also expected the term $:G W^{(\frac{3}{2})}:$, but this is fermionic and hence cannot appear on the right-hand-side.} Again, this completes Step-2.

As regards Step-3, the coefficients in front of $K$, $L$, etc.\ are again fixed by associativity with the bosonic ${\cal N}=4$ generators.
Furthermore, the same is true for the coefficient in front of $\partial W^{(2)}$ and $:KW^{(2)}:$ in the second line. Note that we have not imposed associativity with respect to the supercurrents $G$; if we had, then the coefficient $C_{2,2}^{GG}$ would have also been determined in terms of $n_2$, and similarly $C_{2,2}^{2\uparrow}$ would have been determined in terms of $C_{2,2}^{2}$.

\subsubsection*{The $W^{(\frac{3}{2})}\times W^{(\frac{5}{2})}$ OPE}

The most general ansatz for the $W^{(\frac{3}{2})}\times W^{(\frac{5}{2})}$ OPE is
\begin{align}
W^{(\frac{3}{2})} \times W^{(\frac{5}{2})} \sim & \oplus C_{\frac{3}{2} ,\frac{5}{2}}^{2} \Bigl[ W^{(2)} + \partial W^{(2)} +  :KW^{(2)}: \Bigr] \oplus C_{\frac{3}{2} ,\frac{5}{2}}^{3 }  W^{(3)}   \\
&  \oplus C_{\frac{3}{2}, \frac{5}{2}}^{2\uparrow}  W^{(2)\uparrow} \oplus C_{\frac{3}{2} ,\frac{5}{2}}^{GG} :GG:  \oplus \, 
C_{\frac{3}{2} ,\frac{5}{2}}^{:\frac{3}{2} \frac{3}{2}:} : W^{(\frac{3}{2})} W^{(\frac{3}{2})}: \ . \nonumber
\end{align}
Step-3 determines again the coefficients in front of $\partial W^{(2)}$ and $:KW^{(2)}:$, but not the other coefficients. 

\subsubsection*{The $W^{(\frac{3}{2})} \times W^{(\frac{3}{2})\uparrow}$ OPE}

The most general ansatz for the $W^{(\frac{3}{2})} \times W^{(\frac{3}{2})\uparrow}$ OPE is
\begin{align}
W^{(\frac{3}{2})} \times W^{(\frac{3}{2})\uparrow} \sim & \, C_{\frac{3}{2} ,\frac{3}{2}\uparrow}^{0} \, \Bigl[  K + L + \partial K + :KK: + :KKK: + :KL: + :K \partial K: + \partial L\Bigr] \nonumber  \\
& \oplus C_{\frac{3}{2}, \frac{3}{2}\uparrow}^{2} \Bigl[ W^{(2)} + \partial W^{(2)} +  :KW^{(2)}: \Bigr] \oplus C_{\frac{3}{2}, \frac{3}{2}\uparrow}^{3 }  W^{(3)}  \\
&  \oplus C_{\frac{3}{2}, \frac{3}{2}\uparrow}^{2\uparrow}  W^{(2)\uparrow} \oplus C_{\frac{3}{2} ,\frac{3}{2}\uparrow}^{GG} :GG:  \oplus \, 
C_{\frac{3}{2}, \frac{3}{2}\uparrow}^{:\frac{3}{2} \frac{3}{2}:} : W^{(\frac{3}{2})} W^{(\frac{3}{2})}:  \ .\nonumber 
\end{align}
Associativity with the bosonic ${\cal N}=4$ generators (i.e.\ Step-3) fixes the coefficients in each bracket, but does not constraint the other coefficients that appear in this ansatz.

\subsubsection*{The associativity of (\ref{secondassoc})}

With these preparations we can then study the associativity of (\ref{secondassoc}), and quite remarkably, this already gives rise to many constraints for the structure constants. In particular, assuming that the normalization coefficients $n_s$ are non-zero, we find that 
\begin{align}
\Bigl( C_{\frac{3}{2} ,\frac{3}{2}}^{\, 2} \Bigr)^2 & = - 18 \sqrt{5}\, \frac{(c+24) (c-3)}{c (c+9)(c-6)} \frac{(n_{\frac{3}{2}})^2}{n_2} \ , \label{rel1} \\
\Bigl( C_{2,2}^{\, 2} \Bigr)^2 & = - \frac{112}{\sqrt{5}}\, \frac{(c^2 + 54 c - 90)^2}{c (c+9) (c+24) (c-3) (c-6)} \, n_2\ , \label{rel2} \\ 
C_{\frac{3}{2} ,\frac{5}{2}}^{\, 2} \cdot C_{\frac{3}{2} ,2}^{\frac{5}{2}} & = - 21 \sqrt{\frac{15}{2}} \, \frac{(c+45)(c-6)}{c (c+9) (c-18)} n_{\frac{3}{2}}  \ ,  \label{rel3}\\
C_{\frac{3}{2} ,\frac{3}{2}\uparrow}^{\, 2} \cdot C_{\frac{3}{2} ,2}^{\frac{3}{2}\uparrow} & = 3 \sqrt{\frac{5}{2}} \, \frac{(c-3)(c-6)}{c (c+9) (c+30)} n_{\frac{3}{2}}  \ . \label{rel4}
\end{align}
In addition, we have for example the relation 
\be\label{rel5}
C_{\frac{3}{2}, \frac{5}{2}}^{\, 3}  C_{\frac{3}{2} ,2}^{\frac{5}{2}}   = - i \frac{5^{\frac{3}{4}} \sqrt{21}}{2} 
\sqrt{\frac{ (c+24)(c-3) }{c(c+9)(c-6)}} \, 
\frac{n_{\frac{3}{2}}}{\sqrt{n_2}} \, C_{2,2}^{3}  \ .
\ee
We should stress that the relations (\ref{rel1}) -- (\ref{rel4}) are quite remarkable since they fix the first few structure constants uniquely in terms of the normalization constants $n_{s}$, as well as $c$. This is in marked difference to the situation for the large ${\cal N}=4$ superconformal ${\cal W}_\infty$ algebra, see \cite{Beccaria:2014jra}. In particular, this therefore suggests strongly that the VOA $\mathbb{V}$ does not have any additional free parameters beyond the central charge.

\section{Truncation patterns}\label{sec:Vtrunc}

The constraints on the low-level OPE coefficients, see \eqref{C32322} and \eqref{rel1} -- \eqref{rel5}, are in terms of rational functions of the central charge $c$, and the zeros and poles of these rational functions reflect the truncation patterns of the VOA $\mathbb{V}$.

\subsection{The chiral algebra of $\mathcal{N}=4$ SU$(N)$ SYM }

Recall from the introduction, see in particular eq.~(\ref{cN}), that for the central charge $c=-3 (N^2-1)$ with $N\in \mathbb{N}_{\geq 2}$,  the VOA should truncate to one 
generated by
\begin{equation}
\mathbb{V} = \bigl( {\cal N}=4 \bigr)  \oplus  R^{(\frac{3}{2})} 
\oplus  R^{(2)} 
\oplus \dots
\oplus  R^{(\frac{N}{2})} 
\ ,   
\end{equation}
see also \cite{Beem:2017ooy}.
We can now check whether this is compatible with our answer above, at least for $N=2,3,4$.

\begin{itemize}
\item For $N=2$, we expect that
\begin{equation}\label{eq:ExN2}
N=2: \qquad  \mathbb{V}=({\cal N}=4    )\,,
\qquad \textrm{with} \quad
c = - 3 (2^2-1) = - 9\,.
\end{equation}
We first observe the $c=-9$ appears as a pole on the right-hand-side of the relations \eqref{C32322} and \eqref{rel1} -- \eqref{rel5}. This implies that, at $c=-9$, we are forced to set the normalization factors $n_{\frac{3}{2}}=n_{2}=0$ to zero, and hence that the algebra truncates to the small ${\cal N}=4$ superconformal algebra itself, consistent with the expectation \eqref{eq:ExN2}, see also the discussion below \eqref{C32322}.

\item For $N=3$, we expect 
\begin{equation}\label{eq:ExN3}
N=3: \qquad  {\cal N}=4 \oplus R^{(\frac{3}{2})}\,,
\qquad \textrm{with} \quad
c = - 3 (3^2-1) = - 24\ .
\end{equation}
We note that $c=-24$ is a zero of \eqref{rel1}, which corresponds to the square of the three-point function
\be
\langle  W^{(\frac{3}{2})} \, W^{(\frac{3}{2})} \ W^{(2)}  \rangle \ .
\ee
This means that the $W^{(2)}$ field is not produced in the OPE of $W^{(\frac{3}{2})} \times W^{(\frac{3}{2})}$, i.e.\ that 
\be
C_{\frac{3}{2} ,\frac{3}{2}}^{\, 2} = 0 \ . 
\ee
Furthermore, $c=-24$ is also a pole of \eqref{rel2}, which forces the normalization $n_2=0$.
Both of these are consistent with the truncated spectrum in \eqref{eq:ExN3}.
\item For $N=4$, we expect 
\begin{equation}\label{eq:ExN4}
N=4: \qquad  {\cal N}=4 \oplus R^{(\frac{3}{2})}\oplus R^{(2)}\,,
\qquad \textrm{with} \quad
c = - 3 (4^2-1) = - 45\ .
\end{equation}
We observe that $c=-45$ is a zero of \eqref{rel3}, which
is proportional to the $3$-point function\footnote{The other solution is $c=6$, but for $c=6$ already  eq.~(\ref{C32322}) has a pole, suggesting that for $c=6$ we need to restrict to the ${\cal N}=4$ algebra (just as for $c=-9$).}
\be
\langle W^{(2)} \ W^{(\frac{3}{2})} \, W^{(\frac{5}{2})} \rangle \cdot \langle W^{(\frac{5}{2})}   \ W^{(\frac{3}{2})} \, W^{(2)} \rangle \ .
\ee
This means that the $W^{(\frac{5}{2})}$ field should be absent, 
consistent with the truncated spectrum in \eqref{eq:ExN4}.
\end{itemize}

Our analysis is therefore compatible with the constructions of \cite{Beem:2013sza,Beem:2017ooy,Bonetti:2018fqz,Arakawa:2023cki}. In particular, this furnishes a highly non-trivial consistency check on our quite complicated calculations from above. At the same time, it demonstrates that the VOA $\mathbb{V}$ exists for generic values of $c$, but seems to be otherwise unique. 

\subsection{Systematic truncation analysis}

We can also analyse the truncation patterns more directly by repeating the analysis of Sections~\ref{sec:M92} and \ref{sec:M5}, but with the restriction that the general ansatz for our OPEs only involves the fields in question.
This is to say, when we write down the ans\"atze for the various OPEs, we already take into account the truncation of the spectrum, and then solve for the constraints from associativity with this restricted ansatz. For example, for the algebra that only consists of 
\be\label{4.8}
({\cal N}=4) \oplus R^{(\frac{3}{2})}
\ee
the ansatz for the OPE in (\ref{3232}) is modified to 
\begin{align}\label{3232t}
W^{(\frac{3}{2})} \times W^{(\frac{3}{2})} \sim & \, n_{\frac{3}{2}} \Bigl[ {\bf 1} + c_1 K + c_2 L + c_3 \partial K + c_4 :KK: \Bigr] \ , 
\end{align}
i.e.\ the term proportional to $C_{\frac{3}{2} ,\frac{3}{2}}^{2} W^{(2)}$ is removed. Then one can run the associativity analysis again and check for consistency. We have found that this spectrum only leads to a consistent VOA provided that
\be
c = - 24 \qquad \hbox{or} \qquad c=3 \ . 
\ee

The second solution is subtle since, on the nose, the Jacobi identities are not satisfied. However, the contributions that seem to break the Jacobi identities are in fact null at $c=3$; this then reproduces the results of \cite[see eq.~(1.12), case (A) and (C)]{Bonetti:2018fqz}.\footnote{We thank Leonardo Rastelli for a useful conversation about this issue. Note that their eq.~(1.13) (a) contains a typo: it should be $c=-45$, not $c=-36$.}
Similarly, the VOA 
\be\label{4.11}
({\cal N}=4) \oplus R^{(\frac{3}{2})} \oplus R^{(2)}
\ee
is only consistent at $c=-45$. In particular, the other zero of (\ref{rel3}) does not lead to a consistent algebra since at $c=6$ it is not possible to add the $R^{(\frac{3}{2})}$ multiplet because Step-3 for $s_1=s_2=\frac{3}{2}$ does not have a non-trivial solution.\footnote{One might have guessed this already from the pole in (\ref{c4}), but one can also check this directly.} 

In particular, therefore, we do not see any of the truncation patterns for $c=6M$ with integer $M\geq 2$. Obviously, for $M=1$, it is consistent to restrict just to the ${\cal N}=4$ algebra, but for $M=2$, i.e.\ $c=12$, we have not found any consistent truncation. (In particular, we have checked directly that neither of the spectra of eqs.~(\ref{4.8}) or (\ref{4.11}) are consistent at $c=12$, even taking into account that the Jacobi identities may only hold up to null-fields, as in the $c=3$ case above.) 

\subsection{Low spin modifications}\label{low}

Our analysis relies on the specific form of the assumed spectrum of the VOA $\mathbb{V}$, but one might wonder whether the results would change if we modify the spectrum for small spins. 

The lowest ${\cal N}=4$ multiplet for the VOA $\mathbb{V}$ appears at $s=\frac{3}{2}$, and this is the appropriate set-up for the chiral algebra corresponding to ${\rm SU}(N)$. However, if we were to consider ${\rm U}(N)$ instead, we would get an additional $s=\frac{1}{2}$ multiplet, and correspondingly the special values of the central charge should be 
\be\label{UNc}
c = - 3 N^2 \ . 
\ee

On the other hand, from the viewpoint of the symmetric orbifold algebra, the spectrum for low values of the spin $s$ is a bit uncertain since one may also argue that we should retain the multiplets with $h=j=\frac{w-1}{2}$ (instead of those with $h=j=\frac{w+1}{2}$), see eq.~(\ref{symorbspec}); then the algebra would have an additional multiplet with $s=\frac{1}{2}$ and $s=1$. 

\subsubsection*{Decoupling the spin-$\frac{1}{2}$-multiplet}\label{spin12}

The discussion for the spin $s=\frac{1}{2}$ is relatively easy, and follows essentially the same arguments as in \cite{Goddard:1988wv}.  

Let us start from any VOA that includes a spin-$\frac{1}{2}$-multiplet 
\be
\mathbb{V}_0 \equiv ({\cal N}=4) \oplus R^{(\frac{1}{2})} \oplus \bigoplus_{s\geq 1} N_s \cdot R^{(s)}  \ , 
\ee
where $N_s$ denotes possible multiplicities.
We have checked that we can always ``factor" out the $R^{(\frac{1}{2})}$ multiplet, i.e.\ we can redefine the ${\cal N}=4$ and the $R^{(s)}$ generators with $s\geq 1$, so that they are primary with respect to the fields in $R^{(\frac{1}{2})}$, i.e.\ have a regular OPE. 
This shows that $\mathbb{V}_0$ is isomorphic to two completely decoupled VOAs:
\begin{equation}\label{4.14}
\mathbb{V}_0= \mathbb{V}[\tfrac{1}{2}] \ \oplus \  \mathbb{V}_1 \ , 
\end{equation}
where $\mathbb{V}[\tfrac{1}{2}]$ is the VOA that is generated by the multiplet $R^{(\frac{1}{2})}$. Note that the lowest fields of the multiplet $R^{(\frac{1}{2})}$ simply satisfy the OPE 
\be
W^{(\frac{1}{2})}_r(z) \, W^{(\frac{1}{2})}_s(w)  \sim n_{\frac{1}{2}} \frac{\epsilon_{rs}}{(z-w)} \ , 
\ee
where $r,s\in\{\pm\}$, and $\epsilon_{rs} = - \epsilon_{sr}$ is anti-symmetric.\footnote{This is already the most general ansatz; the coefficient $n_{\frac{1}{2}}$ is just a normalization constant.} In addition to these two free $h=\frac{1}{2}$ fields (which are bosons), the spin-$\frac{1}{2}$ multiplet also contains two (free) $h=1$ fields which are fermions and singlets with respect to the $\mathfrak{su}(2)$ algebra. The bilinears of these four fields then generate an ${\cal N}=4$ superconformal algebra at $c=-3$.  In particular, the ${\cal N}=4$ superconformal algebra that appears in $ \mathbb{V}_1$ of eq.~(\ref{4.14}) is modified by the bilinears of $R^{(\frac{1}{2})}$, and this shifts the central charge as $c\mapsto c+3$.
This is then compatible with the fact that $c_N$ in (\ref{cN}) is shifted by $+3$ relative to $c_N$ in (\ref{UNc}).

On the other hand, the remaining generators that have regular OPEs with the fields from the spin-$\frac{1}{2}$ multiplet, generate a VOA with spectrum $\mathbb{V}_1$, 
\be\label{VOA1}
\mathbb{V}_1 \equiv ({\cal N}=4)  \oplus \bigoplus_{s\geq 1} N_s R^{(s)}  \ .
\ee
Here the ${\cal N}=4$ superconformal generators have been redefined by the bilinears from $W^{(\frac{1}{2})}$, and this shifts $c\mapsto c+3$.

\subsubsection*{The spin-$1$ multiplet}\label{spin1}

The discussion for the spin-$1$ multiplet is a bit more complicated, and this is due to the fact that its structure agrees basically with that of the ${\cal N}=4$ algebra itself. 
As a consequence, one can always add an $s=1$ multiplet and shift the central charge of the overall ${\cal N}=4$ algebra by an arbitrary amount. 

To see this, suppose $\mathbb{V}_1$ denotes an arbitrary ${\cal N}=4$ VOA of the form (\ref{VOA1}) whose ${\cal N}=4$ algebra has generators $K^a$, $G^{\alpha\beta}$, and $L$ and central charge $c$. 
Consider a second ${\cal N}=4$ algebra with generators $\hat{K}^a$, $\hat{G}^{\alpha\beta}$, and $\hat{L}$ and central charge $\hat{c}$ that has regular OPEs with the generators in $\mathbb{V}_1$. 
Then we can define a new VOA $\mathbb{V}_2$ that is generated by 
\be
\mathbb{V}_2 \equiv ({\cal N}=4)_{(c+\hat{c})} \,  \oplus R^{(1)} \oplus \bigoplus_{s\geq 1} N_s R^{(s)}  
\ee
as follows. First of all, the generators of the new ${\cal N}=4$ superconformal algebra are simply 
\be\label{newN4}
K^a + \hat{K}^a \ , \qquad G^{\alpha\beta} + \hat{G}^{\alpha\beta}  \ , \qquad L + \hat{L} \ , 
\ee
and they generate a ${\cal N}=4$ superconformal algebra with central charge $c+\hat{c}$. The original fields in $\mathbb{V}_1$ still transform correctly with respect to these modified ${\cal N}=4$ generators. In addition, we have the new fields 
\be
(\hat{c} \, K^a - c \, \hat{K}^a) \ , \qquad (\hat{c} \, G^{\alpha\beta} -c \, \hat{G}^{\alpha\beta})  \ , \qquad (\hat{c} \, L - c \, \hat{L}) \ , 
\ee
that are, by construction, primary with respect to the ${\cal N}=4$ generators in (\ref{newN4}), and hence generate the new spin-$1$ multiplet.

As a consequence we can always add an arbitrary number of spin-$1$ multiplets to any consistent ${\cal N}=4$ VOA without modifying its consistency; in the process we can shift the central charge by an \textit{arbitrary} amount. The only algebra for which there can therefore be an interesting truncation pattern is the VOA without any spin-$1$ field, i.e. the VOA $\mathbb{V}$ of eq.~(\ref{VOA}) that we have studied in detail in this paper.

\section{Conclusions}\label{sec:concl}

In this paper we have studied the structure of the chiral algebra $\mathbb{V}$ associated to ${\rm SU}(N)$ (or ${\rm U}(N)$) SYM in 4D abstractly. 
At least up to the level to which we have performed our analysis, we have shown that the algebra exists for generic values of the central charge $c$, but we have not found any signs of additional parameters characterizing $\mathbb{V}$ beyond the central charge.\footnote{Note that this rigidity of the $\mathcal{N}=4$ chiral algebra can provide a proof of the twisted holography of \cite{Costello:2018zrm}; we thank Kevin Costello and Davide Gaiotto for pointing this out to us.} 
We have furthermore analysed for which values of the central charge $c$ the algebra truncates to one that is finitely generated by the first $N$ multiplets, and we have found strong evidence that the only consistent solutions are\footnote{For $N=3$, also $c=3$ is possible, but there is no second solution for $N=4$, see eq.~(\ref{4.11}) above. 
The same comment also applies to the ${\rm U}(N)$ spectrum in eq.~(\ref{UNcase}).} 
\be
\mathbb{V}_N = ({\cal N}=4) \oplus\ \bigoplus_{s=\frac{3}{2}}^{\frac{N}{2}} R^{(s)} \qquad \hbox{only for $c=-3(N^2-1)$.}
\ee

Furthermore, if we add to $\mathbb{V}$ a spin-$\frac{1}{2}$ multiplet, the VOA has the correct spectrum to correspond to ${\rm U}(N)$ SYM in 4D. As discussed in Section~\ref{spin12} it has essentially the same truncation pattern as $\mathbb{V}$, except that the central charge is shifted by $-3$, i.e.\ the algebra truncates to 
\be\label{UNcase}
\mathbb{V}^{(1)}_N = ({\cal N}=4) \oplus\ R^{(\frac{1}{2})} \oplus \bigoplus_{s=\frac{3}{2}}^{\frac{N}{2}} R^{(s)} \qquad \hbox{only for $c=-3N^2$.}
\ee
Again, this reproduces what one expects from the 4D perspective. 

\bigskip

Both these truncation patterns fit beautifully with the 4D SYM expectations.
However, the absence of any truncation at $c=6M$ for integer $M\geq 2$ seems to be at odds with what the twisted holography analysis of \cite{Costello:2020jbh} suggests.
In the remainder of this section we will therefore discuss possible explanations of this finding.
\medskip

In relating the chiral algebra of \cite{Costello:2020jbh} to $\mathbb{V}$, we have made the assumption that one can ``decouple" the $\mathbb{T}^4$, i.e.\ remove the tensor factor coming from the cohomology of $\mathbb{T}^4$, and in the supergravity limit, this seems to be very reasonable. This decoupling is also suggested by the worldsheet analysis of \cite{Li:2023tyx} for which $\mathbb{V}$ arises from the  $\mathbb{T}^4$-independent part of string theory on ${\rm AdS}_3 \times {\rm S}^3 \times \mathbb{T}^4$ with pure R-R flux. Since this is a perturbative analysis, it also assumes large $M$. However, it is conceivable that this decoupling breaks down at finite $M$, i.e.\ that it is the finite $M$ (or finite $c$) properties of $\mathbb{V}$ that are not compatible with the symmetric orbifold expectations. Finite $M$ properties of the symmetric orbifold were also recently studied in \cite{LL}. 

Relatedly, since the twisted holography answer differs slightly from the BPS supergravity analysis of \cite{deBoer:1998kjm} for small spins,\footnote{This is not unexpected, see a similar discussion in \cite{deBoer:1998us}.} it is conceivable that the algebra that appears from the ${\rm AdS}_3 \times {\rm S}^3$ perspective differs slightly from the ansatz we have considered, e.g.\ we may have to add an additional $s=\frac{1}{2}$ and/or $s=1$ multiplet to the spectrum of $\mathbb{V}$ to find a consistent algebra, see also the discussion at the beginning of Section~\ref{low}. 
As we have explained in Section~\ref{spin12}, the first option does not modify the answer significantly since we can always factor out any spin-$\frac{1}{2}$ multiplet, and this only shifts the central charge by $c\mapsto c+3$. 
On the other hand, we cannot rule out the second option, i.e.\ adding a spin $s=1$ multiplet, since this is always possible (unless one makes further assumptions about how the spin-$1$ multiplet interacts with the other generators). However, since this would allow one to shift the central charge by an \textit{arbitrary} amount, see Section~\ref{spin1}, this scenario would lose any rigidity and predictive power.

And finally, while the twisted holography analysis makes it very plausible that there should be a chiral algebra associated to the symmetric orbifold, a direct construction has, to our knowledge, not been given to this date. 
It is therefore also conceivable that there esists a more subtle obstruction. 
It would therefore be very interesting to construct this chiral algebra from first principles. 
\bigskip

\noindent{\bf Acknowledgments:} We thank Rajesh Gopakumar and Edward Mazenc for discussions about related topics, and Lorenz Eberhardt, Ji Hoon Lee, Beat Nairz, and Leonardo Rastelli for useful conversations and correspondences. 
The work of MRG is supported by a personal grant from the Swiss National Science Foundation, and the work of his group at ETH is also supported in part by the Simons Foundation grant 994306  (Simons Collaboration on Confinement and QCD Strings), as well as the NCCR SwissMAP that is also funded by the Swiss National Science Foundation. 
The work of WL is supported by NSFC  grants No.\  12275334, 12447101, and 12247103. 
We are also grateful to KITP for hospitality during the programme ``What is string theory? Weaving perspectives together"  when part of this work was carried out.

\appendix

\section{${\cal N}=4$ conventions} \label{app:N4}

\subsection{${\cal N}=4$ superconformal algebra}

We work with the standard ${\cal N}=4$ conventions. 
The fields of the ${\cal N}=4$ superconformal algebra consist of the stress energy tensor $L(z)$ with central charge $c$, $\mathfrak{su}(2)$ currents of level $k=\frac{c}{6}$, for which we choose the normalizations
\be
\begin{array}{rclrcl}
K^3(z) K^3(w) &\sim & {\displaystyle \frac{k}{2 (z-w)^2} }\ , \qquad & K^+(z) K^-(w) & \sim & {\displaystyle \frac{k}{(z-w)^2} + \frac{2 K^3(w)}{(z-w)} }\\[10pt]
K^3(z) K^\pm (w) & \sim & {\displaystyle \pm \frac {K^\pm(w)}{(z-w)}} \ , \qquad & & & 
\end{array}
\ee
as well as supercharges $G^{\alpha\beta}$ with $\alpha,\beta\in\{\pm\}$ of spin $h=\frac{3}{2}$, with OPEs 
\be
K^a(z) G^{\alpha\beta}(w) \sim \frac{D^{(1/2)}(t^a)^{\alpha}{}_{\gamma}  \, G^{\gamma\beta}(w)}{(z-w)} \ , 
\ee
where the non-trivial matrix elements are 
\be\label{me1}
D^{(1/2)}(t^+)^{-}{}_{+} = D^{(1/2)}(t^-)^{+}{}_{-}  = 1 \ , \qquad D^{(1/2)}(t^3)^+{}_{+} = \tfrac{1}{2} = - D^{(1/2)}(t^3)^-{}_{-}  \ . 
\ee
Furthermore, we have 
\be
G^{\alpha\beta}(z) \, G^{\gamma\delta}(w) \sim - \frac{\frac{4c}{3} \epsilon^{\alpha\gamma} \epsilon^{\beta\delta}}{(z-w)^3}  - \frac{8 \epsilon^{\beta\delta} D^{\alpha\gamma}(t_a) K^a(w)}{(z-w)^2} - 
\frac{4 \epsilon^{\alpha\gamma} \epsilon^{\beta\delta} L(w) + 4 \epsilon^{\beta\delta} D^{\alpha\gamma}(t_a) \partial K^a(w)}{(z-w)} \ ,
\ee
where the non-trivial matrix elements are 
\be
D^{+-}(t_3) = 1 =  D^{-+}(t_3) \ , \qquad D^{++}(t_+) = - 1  \ , \qquad D^{--}(t_-) =  1 \ . 
\ee
(They are obtained from the elements in (\ref{me1}) upon raising and lowering the spinor indices with the $\epsilon$ tensor, $\epsilon^{+-} = +1$ and $\epsilon^{-+}=-1$, while raising and lowering the vector indices with the Cartan metric.)

\subsection{The ${\cal N}=4$ BPS representations}\label{app:A.1}

In this subsection we explain how the ${\cal N}=4$ superconformal algebra generators act on the fields in the BPS multiplet $R^{(s)}$. We shall work with the conventions that the latter fields are denoted by 
\be
\begin{array}{lll}
W^{(s)}_{m}  \qquad & m=-s,\ldots,s  \qquad \qquad \qquad & (h=s)\\
W^{(s)[\alpha]}_{m}  \qquad & m=-(s-\tfrac{1}{2}),\ldots,(s-\tfrac{1}{2})  \qquad \qquad \qquad & (h=s+\tfrac{1}{2})\\
W^{(s)\uparrow}_{m}  \qquad & m=-(s-1), \ldots, (s-1)   & (h=s+1) \ , 
\end{array}
\ee
where $\alpha\in\{ \pm\}$. Furthermore we take the matrix elements of the $\mathfrak{su}(2)$ representations to be given by 
\begin{align}
D^{(j)}(t^3)_{n,m} & = n\, \delta_{n,m} \nonumber \\
D^{(j)}(t^+)_{n,m} & = \sqrt{j(j+1) - n(n+1)}\, \delta_{m,n+1}  \label{Dmatrix} \\ 
D^{(j)}(t^-)_{n,m} & = \sqrt{j(j+1) - n(n-1)}\, \delta_{m,n-1} \ . \nonumber 
\end{align}
Then the OPEs of the $\mathfrak{su}(2)$ currents and the stress tensor are of the form 
\begin{align}
K^a(z) W^{(s)}_m(w) & \sim \frac{D^{(s)}(t^a)_{mm'} \, W^{(s)}_{m'}(w)}{(z-w)} \\
L(z) W^{(s)}_m(w) & \sim \frac{s W^{(s)}_m(w)}{(z-w)^2}  + \frac{\partial W^{(s)}_m(w)}{(z-w)}  \\ 
K^a(z) W^{(s)[\alpha]}_r(w) & \sim \frac{D^{(s-1/2)}(t^a)_{rr'} W^{(s)[\alpha]}_{r'}(w)}{(z-w)} \\ 
L(z) W^{(s)[\alpha]}_r(w) & \sim \frac{(s+\frac{1}{2}) W^{(s)[\alpha]}_r(w)}{(z-w)^2}  + \frac{\partial W^{(s)[\alpha]}_r(w)}{(z-w)}  \\ 
K^a(z) W^{(s)\uparrow}_n(w) & \sim \frac{D^{(s-1)}(t^a)_{nn'} W^{(s)\uparrow}_{n'}(w)}{(z-w)} \\ 
L(z) W^{(s)\uparrow}_n(w) & \sim \frac{(s+1) W^{(s)\uparrow}_n(w)}{(z-w)^2}  + \frac{\partial W^{(s)\uparrow}_n(w)}{(z-w)}  \ . 
\end{align}
For the action of the supercharges we find by self-consistency (i.e.\ associativity of the OPE) that they take the form 
\begin{align}
G^{\alpha\beta}(z) \, W^{(s)}_m(w) & \sim \frac{\rho^{(s,-)}_{\alpha,m} W^{(s)[\beta]}_{m+ \frac{\alpha}{2}}(w)}{(z-w)} \ , \\
G^{\alpha\beta}(z) \,  W^{(s)[\gamma]}_r(w) & \sim \epsilon^{\beta\gamma} \Biggl[ c_1(s) \frac{{\rho}^{(s-1/2,+)}_{\alpha,r} W^{(s)}_{r+ \frac{\alpha}{2}}(w)}{(z-w)^2} \\
&  \quad \qquad + \frac{\rho^{(s-1/2,-)}_{\alpha,r} W^{(s)\uparrow}_{r+ \frac{\alpha}{2}}(w) + c_3(s)\,  {\rho}^{(s-1/2,+)}_{\alpha,r}  \partial W^{(s)}_{r+ \frac{\alpha}{2}}(w) }{(z-w)} \\
& \quad \qquad + \frac{c_4(s) \, C^{(s-\frac{1}{2},-)}(a,m;\alpha,r)  :K^aW^{(s)}_{m}: (w)} {(z-w)}  \Biggr] \ ,\\
G^{\alpha\beta}(z) \, W^{(s)\uparrow}_{n}(w) & \sim c_5(s) \, \frac{{\rho}^{(s-1,+)}_{\alpha,n} W^{(s)[\beta]}_{n+ \frac{\alpha}{2}}(w)}{(z-w)^2} \\ 
& + \frac{c_6(s)\,  \bigl({C}^{(s-1,-)}(a,r;\alpha,n) + A_s {C}^{(s-1,+)}(a,r;\alpha,n) \bigr) :K^aW^{(s)[\beta]}_{r}: (w)} {(z-w)} \\
& + \frac{c_7(s)\,  {\rho}^{(s-1,+)}_{\alpha,n} \partial W^{(s)[\beta](w)}_{n+\frac{\alpha}{2}} 
+ c_8(s) D^{(s)}(\gamma,m; \alpha,n) :G^{\gamma \beta} W^{(s)}_m: }{(z-w)} \ , 
\end{align}
where $\rho^{(j,\pm)}_{\alpha,m}$ is the Clebsch-Gordan coefficient corresponding to the tensor product decomposition,
\be
j \otimes \tfrac{1}{2}\rightarrow (j\pm\tfrac{1}{2}): \qquad 
\rho^{(j,\pm)}_{\alpha,m}= \langle j\pm\tfrac{1}{2},m + \tfrac{\alpha}{2} |\!| j,m;\tfrac{1}{2} , \tfrac{\alpha}{2} \rangle \ .
\ee
Furthermore, $C^{(j,\pm)}(a,m;\alpha, r)$ describes the coupling between 
\be
 |j ,r\rangle  \otimes |\tfrac{1}{2},\tfrac{\alpha}{2}\rangle \ \stackrel{j\pm \frac{1}{2}} {\longleftrightarrow } \ \  |j+\tfrac{1}{2},m\rangle \otimes |1,a\rangle  \ , 
\ee
i.e. 
\begin{align}
C^{(j,\pm)}(a,m; \alpha,r)  = & \delta\bigl(m + a - (r+\tfrac{\alpha}{2})\bigr) \times \\
& \  \times  \epsilon_a 
\langle j\pm \tfrac{1}{2},m+a |\!|  (j+\tfrac{1}{2}),m;1,a \rangle \cdot 
\langle  j\pm \tfrac{1}{2},\tfrac{\alpha}{2} + r|\!|  j,r; \tfrac{1}{2} , \tfrac{\alpha}{2} \rangle \ , \nonumber 
\end{align}
where the factors of $\epsilon_a$ appear since the $\mathfrak{su}(2)$ action on the currents $K^a$ via commutators differs from the action on the  $|1,a\rangle$ states as defined via (\ref{Dmatrix}) by the rescaling 
\be\label{epsd}
\epsilon_+ = -1 \ , \qquad \epsilon_0 = \sqrt{2} \ , \qquad \epsilon_- = 1 \ .
\ee
Finally, the $D^{(s)}(\gamma,m; \alpha,n)$ coefficient describes the coupling between 
\be
 |s-1,n\rangle \otimes |\tfrac{1}{2},\tfrac{\alpha}{2}\rangle \ \ \stackrel{s-\frac{1}{2}} {\longleftrightarrow } \ \ 
|s,m\rangle  \otimes |\tfrac{1}{2},\tfrac{\gamma}{2}\rangle \ , 
\ee
i.e.\
\begin{align}
D^{(s)}(\gamma,m; \alpha,n)  = & \delta\bigl(m + \tfrac{\gamma}{2} - (\tfrac{\alpha}{2}+n)\bigr) \times \\
& \  \times 
\langle s- \tfrac{1}{2},n+\tfrac{\alpha}{2} |\!|(s-1),n ;  \tfrac{1}{2},\tfrac{\alpha}{2}  \rangle \cdot 
\langle  s- \tfrac{1}{2},\tfrac{\gamma}{2} + m|\!|  s,m; \tfrac{1}{2} , \tfrac{\gamma}{2} \rangle \ . \nonumber 
\end{align}
For the Clebsch-Gordan coefficients we have used the explicit formula 
\begin{align}
& \langle J,m_1+m_2|\!|  j_1,m_1;j_2,m_2\rangle = \\
& \qquad \sqrt{\frac{(2J+1) \, (J+j_1 - j_2)!\, (J+j_2 - j_1)! \, (j_1 + j_2 - J)!}{(j_1 + j_2 + J +1)!}} \nonumber \\
& \qquad \times \sqrt{(J+M)! \, (J-M)! \, (j_1 + m_1)!\, (j_1 - m_1)!\, \, (j_2 + m_2)!\, (j_2 - m_2)!} \nonumber \\[4pt]
& \qquad \times \sum_{k\geq 0} \Biggl[ \frac{(-1)^k}{k!\, (j_1+j_2-J-k)! \, (j_1 - m_1 -k)! } \nonumber \\[4pt]
& \qquad \qquad \qquad \times \frac{1}{ (j_2+m_2 - k)! \,(J-j_2+m_1+k)! \,(J-j_1-m_2+k)!}  \Biggr]\ ,
\end{align}
where $M=m_1+m_2$, and the sum over $k$ runs over those integers for which the argument of the factorial in the denominator is non-negative. With these conventions, the parameters $A_s$ equal 
\be
A_s =  \frac{1}{\sqrt{s (2s+1)}}  \ ,
\ee
while the $c(s)$ take the values 
\be
c_1(s) = - 4 \sqrt{2s (2s+1)} \ , \qquad 
c_3(s) = - 4 \sqrt{\frac{2s+1}{2s}}  \ , \qquad 
c_4(s) =  - 24 \frac{2s+1}{ c + 12 (s+1)}  \ ,
\ee
as well as 
\be
c_5(s) = \frac{ (2s+1) (c + 12 s + 6)}{c + 12 (s+1)} \sqrt{\frac{8 (2s-1)}{s}}\ , \qquad 
c_6(s) =  \frac{24 (2 s+1)}{c + 12 (s+1)} \ , \qquad
\ee
and
\be
c_7(s) = \sqrt{\frac{8 (2s-1)}{s}} \ , \qquad 
c_8(s) = - \frac{12 (2s +1)}{s( c + 12 (s+1))}  \sqrt{2s (2s -1)} \ .
\ee

\bibliographystyle{JHEP}

\end{document}